\magnification=1200
\def\qed{\unskip\kern 6pt\penalty 500\raise -2pt\hbox
{\vrule\vbox to 10pt{\hrule width 4pt\vfill\hrule}\vrule}}
\centerline{THE ORIGIN OF LIFE SEEN FROM THE POINT OF VIEW}
\centerline{OF NON-EQUILIBRIUM STATISTICAL MECHANICS.}
\bigskip\bigskip\bigskip\bigskip
\centerline{by David Ruelle\footnote{$\dagger$}{Math. Dept., Rutgers University, and 
IHES, 91440 Bures sur Yvette, France. email: ruelle@ihes.fr}.}
\bigskip\bigskip\bigskip\bigskip\noindent
	{\leftskip=2cm\rightskip=2cm\sl Abstract. This note presents a minimal approach to the origin of life, following standard ideas.  We pay special attention to the point of view of non-equilibrium statistical mechanics, and in particular to detailed balance.  As a consequence we propose a characterization of pre-biological states.\par}
\vfill\eject
\noindent
{\bf 1. An outline of the origin of life.}
\medskip
	Life seems to have appeared on Earth soon after this was permitted by the ambient conditions.  It is therefore plausible that the origin of life is a natural consequence of the laws of physics: in some range of values for the ambient parameters, some form of life should appear with high probability.  The laws of physics include conservation of energy and increase of entropy (first and second law of thermodynamics).  What we call life exhibits highly improbable structures, but this does not contradict the overall increase of entropy, as argued for instance by Schr\"odinger [9].  Still, one would like to gain understanding of how and why the improbable structures of life arise: the catalytic system which organizes biological chemistry, the replication system which deals with biological information, the machinery of membranes enclosing cells, etc.  These apparently improbable structures should be outputs, not inputs of a theory of the origin of life.
\medskip
	Non-equilibrium statistical mechanics is the physical theory that should be used to understand the origin of life.  What one calls non-equilibrium statistical mechanics is however only a collection of results of limited applicability.  There is a beautiful theory of non-equilibrium close to equilibrium (fluctuation-dissipation theorem, etc.) but the life processes are typically very far from equilibrium.  There are mathematically profound results based on the theory of hyperbolic dynamical systems (Gallavotti-Cohen fluctuation theorem for entropy production [7], linear response away from equilibrium).  There are also general developments assuming a stochastic modeling of microscopic dynamics (Evans, Jarzynski, Crooks, etc.).  The detailed balance formula (based on time-reversal invariance and assuming an environment at a given temperature) is another tool of non-equilibrium statistical mechanics.  Detailed balance does not depend essentially on stochastic assumptions and makes limited but specific non-equilibrium predictions.
\medskip
	A brief digression on stochastic forces may be appropriate here.  Markovian stochasticity is a natural simplifying approximation for microscopic dynamics.  It can be justified to some extent from deterministic chaotic assumptions on the mechanical laws for time evolution.  Stochastic forces can be used to justify the fluctuation-dissipation formula, the fluctuation formula for entropy production, and the detailed balance formula.  But using as in [7] deterministic rather than stochastic forces is not only more satisfactory from a basic conceptual point of view, it is also likely to lead to specific predictions of greater physical interest.  In particular detailed balance does not depend on Markovian stochasticity in an essential way (see [8]), and predicts useful relations for the chemical reaction rates, while more general formulas based on stochasticity lack predictive value.
\medskip
	In the vast literature on the origin of life there are relatively few contributions based on non-equilibrium statistical mechanics.  Let us mention Andrieu and Gaspard [1], England [5], Fox [6].  England's paper has the interest of being based on the detailed balance formula.
\medskip
	Beginning with a chemical system out of equilibrium, we shall try to understand the origin of life as a necessary process (under suitable general initial conditions) resulting from the laws of statistical mechanics.  We have thus to see how organized metabolism and replication of information spontaneously arise.  As a starting point we may think of a liquid bath (water containing various solutes) interacting with some pre-metabolic systems.  The pre-metabolic systems are chemical associations which may be carried by particles floating in the liquid, or contained in cracks of a solid boundary of the liquid, etc.  We shall want that local equilibration times for temperature and pressure, as well as diffusion times for solutes in the liquid, are small compared with the times over which we observe the evolution of the pre-metabolic systems (these are natural conditions, needed in particular for detailed balance).  Apart from that we shall attempt at using only basic physical principles, and shall avoid deriving central life processes from Markovian stochasticity assumptions.
\medskip
	In what follows we shall not propose a specific scenario for the origin of life.  Instead of this we shall outline a few plausible steps leading to more and more complex states of pre-metabolic systems so that something like life may naturally arise.  What actually took place may have been different, and was certainly very complicated.  But the point is that there are plausible scenarios which lead necessarily to life, or something like it.
\medskip	
	 We shall consider situations related to different time scales:
\medskip
	(a) Slowly evolving pre-metabolic states.
\medskip
	The pre-metabolic system is in a stationary state and interacts with a fixed environment (i.e., fixed temperature, pressure and solute composition of the fluid) or the pre-metabolic state evolves slowly with an imposed slowly evolving environment.  [The ideas of detailed balance theory clarify the microscopic structure of slowly evolving states, see Section 2].
\medskip
	(b) Co-evolution with environment.
\medskip
	On a longer time scale, the activity of pre-metabolic systems may change the parameter values of the environment.  This in turn changes the states of the systems interacting with the bath and some may disappear.
\medskip
	(c) Creation of new states.
\medskip
	Also on a longer time scale, fluctuations in the chemistry of a slowly evolving state of a pre-metabolic system may cause the creation of a new slowly evolving state.  The creation of the new state may for instance involve the creation of new polymers.  The new state may have a more complex pre-metabolic activity than the original one.  Some systems may divide into pieces which evolve differently.  Asymptotically this may lead to the coexistence of possibly several pre-metabolic systems.
\medskip
	(d) Competition of pre-metabolic systems.
\medskip
	In view of (b) and (c) the distribution of steady states of pre-metabolic systems is expected to be time dependent on a sufficiently long time scale.  Some systems may be forced to disappear (as in (b)) others may arise (as in (c)).
\medskip
	(e) Towards biological evolution.
\medskip
	If a pre-metabolic system contains polymers and the growth of one polymer is influenced by the nature of already present polymers, the competition mentioned in (d) operates and results in a selection process: biological evolution has begun.
\medskip
	The purpose of the present note is to attempt a more precise discussion of the above remarks by using basic ideas of non-equilibrium statistical mechanics.  In view of this we have just presented some accepted or acceptable ideas on pre-biological or pre-metabolic systems.  Note that one such system may occupy several distinct regions of space (just as biological species may consist of different individuals).  But pre-metabolic systems have a discrete structure: we are not thinking of a homogeneous pre-biological soup.
\medskip
	The spirit of our approach is related to that of England [5].  We shall however concentrate our attention on almost stationary states rather than replicating systems (which need not be present).  In view of detailed balance the Gibbs free energy production plays an important role in the stationary state approach as it does in England's paper (this is not too astonishing).
\medskip
	For simplicity we shall in what follows speak of systems instead of pre-metabolic systems, and metabolic activity instead of pre-metabolic activity.
\medskip
	Let us now discuss the situations (a)-(e) in more details.
\bigskip\noindent
(a) {\bf Slowly evolving states.}
\medskip
	We consider a system $M$ interacting with a bath containing various solutes.  The system $M$ is initially in a state $J$ which we assume to have a large more or less homogeneous surface interacting metabolically with the bath.  For definiteness we let $J$ describe a unit piece of $M$, say a unit element of metabolically active surface.  We choose a time interval $\tau$ large with respect to local equilibration times.  In this time interval the state $J$ becomes a state $K$ with Gibbs free energy change $\Delta_M=G(K)-G(J)$ while there is a corresponding Gibbs free energy change $\Delta_*$ of the solutes in the bath as they interact with the system $M$.  The second law of thermodynamics gives
$$	\Delta_M+\Delta_*\le0   $$
\indent
	Let the state $K$ become $L$ after another time interval $\tau$:
$$	J\to K\to L\to\cdots   $$
We may say that $K\to L$ is metabolically equivalent to $J\to K$ if the composition of the bath is kept constant and if the metabolically active surface of $K$ is equivalent to the metabolically active surface of $J$ (the bulk of $K$ may be different from the bulk of $J$ so that $\Delta_M\ne0$).  Metabolic equivalence of $J\to K$ and $K\to L$ implies that $\Delta_M$ and $\Delta_*$ are the same for $J\to K$ and $K\to L$, and we may say that $J\to K\to L\to\cdots$ describes an almost stationary state of $M$.
\medskip
	We expect that a small change in the composition of the bath, such that $\delta\Delta_*=\Delta_*(K\to L)-\Delta_*(J\to K)$ need not vanish, will give rise to a small change
$$	\delta\Delta_M=\Delta_M(K\to L)-\Delta_M(J\to K)=\kappa\delta\Delta_*   $$
with proportionality constant $\kappa$.  In this situation we may speak of a slowly evolving state.  Note that $\Delta_*$ corresponds to a channel $\alpha$ with initial state $\alpha^{\rm in}$ and final state $\alpha^{\rm out}$ for the solutes in the bath.  In particular $\Delta_*$ depends on the channel $\alpha$ for $J\to K$.  [The situation where $J\to K$ involves various channels $\alpha$ with different probabilities can be discussed using detailed balance, see Section 2].
\bigskip\noindent
(b) {\bf Co-evolution with the environment.}
\medskip
	Consider now several systems $M_j\,(j=1,\dots)$ in slowly evolving states  $K_j$ interacting with the same bath.  Denote by $\Delta_{*j}$ the change of free energies of the solutes for the time interval $\tau$ and a given $K_j$.  The changes $\Delta_j$ in Gibbs free energies for the states $K_j$ satisfy
$$	-(\Delta_{Mj}+\Delta_{*j})\ge0   $$
\indent
	Let us discuss the situation where the growth of the $K_j$ slowly influences the concentration of solutes in the bath, and the concentration of solutes influences the growth of the $K_j$ in return.  This could lead to a nontrivial time evolution of the bulks of the $K_j$ (described by their Gibbs free energies $G_j$).  If ``nutrients'' are supplied to the bath at a limited rate, one can easily obtain the simple situation where the system tends to a stationary state.  In particular we have asymptotically $\Delta_{Mj}=0$.  We have thus two possibilities:
\medskip
	($\alpha$) $\Delta_{Mj}+\Delta_{*j}$ vanishes so that $\Delta_{*j}$ also vanishes.  This means that the system $M_j$ is {\it dead}, it has no metabolic activity.
\medskip
	($\beta$) $\Delta_{Mj}+\Delta_{*j}$ does not vanish, so that $-\Delta_{*j}>0$.  There is then metabolic activity in the sense that $M_j$ catalyzes the transformation $\alpha^{\rm in}\to\alpha^{\rm out}$ of the solutes.
\medskip
	The simplest case of interest is when there is just one system $M$ in a state $K$ with metabolic activity.
\bigskip\noindent
(c) {\bf Fluctuations and the creation of new stationary states.}
\medskip
	The chemical reactions taking place in a metabolically active stationary (or slowly evolving) state $K$ will undergo microscopic fluctuations.  If such a fluctuation produces a chemical molecule $X$ of a new kind, this molecule may have catalytic properties which will change the chemistry of $K$ locally.  After a while the molecule $X$ will be destroyed, and this will usually put an end to the fluctuation initiated with $X$.  It is however possible that $X$ catalyzes directly or indirectly the production of a new molecule $Y$ which has the autocatalytic property that, before being destroyed, it has led to the production of other copies of the molecule $Y$, more than 1 in the average.  The state $K$ is thus changed into a state $K'$ with a different metabolism.  We may assume that $K'$ is slowly evolving, but it is usually not a stationary state in the given environment.  If the change $\Delta'$ of the Gibbs free energy of $K'$ after time $\tau$ is $<0$ (and remains $<0$) the system $M$ will rapidly disappear.
\medskip
	The interesting situation is when $\Delta_{M'}>0$, i.e., when the new state $K'$ grows with time.  [This corresponds to rare fluctuations of the original state $K$].  In this case the state $K'$ of $M$ and the states $K_1,\ldots$ of $M_1,\ldots$ possibly present in the same bath, will evolve (in the simplest case) towards stationary states $\tilde K,\tilde K_1,\ldots$ as discussed in (b) above.
\medskip
	The process of creation of new stationary states just described is somewhat schematic but conceptually clear.  A state $K$ may also undergo an aging process with the progressive creation of more and more complex polymers.  [An interesting example of system of that sort has been proposed for the origin of life: it evolves from a plausible mixture of polyphosphate and small organic molecules [2]].
\bigskip\noindent
(d) {\bf Competing systems.}
\medskip
	Consider again several systems $M_j\,(j=1,\dots)$ in non-stationary states  $K_j$ interacting with the same bath.  We assume that the co-evolution of the $K_j$ and the solutes in the bath gives asymptotic stationary states $\tilde K_j$ of the $M_j$ and a stationary composition of the bath.  The use of solutes (described by $\alpha_j^{\rm in}\to\alpha_j^{\rm out}$) will in general be different for the different states $\tilde K_j$.  Therefore the systems $M_j$ will not necessarily compete for ``nutrients'' and can coexist in states $\tilde K_j$ corresponding to different metabolisms.
\medskip
	Over a long timescale, some systems $M_j$  may divide spatially into pieces.  The state of some pieces may become new states as in (c) and some pieces may disappear as in (b).  Also, if a system $M_j$ is in a state $K_j$ it may release some molecule with catalytic activity which may reach another system and transform it (horizontal metabolic transfer).  We have thus a picture of competing systems in states with different metabolisms.  In this picture different systems have different spatial localizations, but need not be separated by membranes.  All we need are different systems in slowly evolving states, interacting with a bath, with chemical fluctuations which occasionally cause a new state to arise with greater metabolic complexity.
\medskip
	Metabolic complexity is not in itself a competitive advantage, but may allow growth in a given environment: $\Delta_{Mj}>0$.  Some systems will thus grow for a while, modify the environment, and some other systems will thus be eliminated.  We expect thus that systems with fast metabolism will progressively dominate in an environment which becomes progressively poorer in ``nutrients''.
\bigskip\noindent
(e) {\bf Towards biological evolution.}
\medskip
	Consider a system using catalytic molecules, with rapid and complex metabolism.  In this situation chemical fluctuations as discussed in (c) will occur frequently.  In particular, the creation of (random) polymers will be favored including some with catalytic activity.  When autocatalytic activity arises in a class of polymers, selection can begin to operate within this class, and one can say that biological evolution begins.
\medskip
	We have restricted our ideas on the origin of life to a minimum, compatible with popular views (see [9], [4], [3]).  Rigorous non-equilibrium statistical mechanics makes a limited contribution to the sketch of the origin of life which we have discussed.  In Section 2 we present detailed balance results as they may contribute to an understanding of pre-biological states.  A characterization of pre-biological states is attempted in Section 3.
\bigskip\noindent
{\bf 2. Detailed balance.}
\medskip
	We consider as above a system $M$ immersed in a large bath of fluid (water containing various solutes).  The temperature, the pressure, and the chemical potentials of the solutes are fixed.  We let $J,K$ denote states of the system $M$.  These states are statistical descriptions for the microscopic positions and momenta of the atoms composing $M$ and a small body of fluid around $M$.  For our purposes $J$ and $K$ should be metastable states with lifetimes large compared with local equilibration times for pressure, temperature, and solute concentration in the fluid.  We do not discuss here how these conditions are implemented mathematically (see [8] for details).
\medskip
	Denote by $\pi_\tau(J\to K)$ the conditional probability that, starting from the state $J$, we end up in the state $K$ after time $\tau$.  The transition $J\to K$ may occur through various channels $\alpha$ corresponding to changes $\alpha^{\rm in}\to\alpha^{\rm out}$ of the number of molecules of various kinds which are absorbed or rejected in the bath in the course of $\alpha$.  The solute molecules may be metastable, but we assume that they undergo chemical changes only in contact with the system $M$.  This simplifying assumption just means that the chemical composition of the bath is kept constant.
\medskip
	We may write $\pi_\tau(J\to K)$, $\pi_\tau(K\to J)$ as sums over channels:
$$	\pi_\tau(J\to K)=\sum_\alpha\pi_\tau^\alpha(J\to K)\qquad,
	\qquad\pi_\tau(K\to J)=\sum_\alpha\pi_\tau^{-\alpha}(K\to J)\eqno{(1)}   $$
where $-\alpha$ is the reverse $\alpha^{\rm out}\to\alpha^{\rm in}$ of channel $\alpha$.  We shall use the probabilities $p^\alpha,\bar p^\alpha$ of the channels $\alpha,-\alpha$:
$$	p^\alpha={\pi_\tau^\alpha(J\to K)\over\pi_\tau(J\to K)}\qquad,\qquad
	\bar p^\alpha={\pi_\tau^{-\alpha}(K\to J)\over\pi_\tau(K\to J)}\eqno{(2)}   $$
We shall also use the identity (see [8])
$$	{\pi_\tau^{-\alpha}(K\to J)\over\pi_\tau^\alpha(J\to K)}=\exp[\beta(\Delta_M^\alpha+\Delta_*^\alpha)]\eqno{(3)}   $$
where $\Delta_M^\alpha$ and $\Delta_*^\alpha$ are the changes in Gibbs free energy of the system $M$ and the bath in the transition $(J,\alpha^{\rm in})$ to $(K,\alpha^{\rm out})$.  This identity results from the time reversal invariance of the basic laws of physics.  Note that $\Delta_*^\alpha G$ takes into account the concentration of the solutes (via their chemical potentials).  From (1),(2),(3) we obtain the following generalized detailed balance relation (see [8], equation (1.6)):
$$	{\pi_\tau(K\to J)\over\pi_\tau(J\to K)}
	=\sum_\alpha p^\alpha\exp[\beta(\Delta_M^\alpha+\Delta_*^\alpha)]\eqno{(4)}   $$
[In formula (4) the change in Gibbs free energy in the transition $(J,\alpha^{\rm in})\to(K,\alpha^{\rm out})$ is $\Delta_M^\alpha=-\beta\Delta_M^\alpha S+\Delta_M^\alpha H$ for the system $M$ and $\Delta_*^\alpha=-\beta\Delta_*^\alpha S+\Delta_*^\alpha H$ for the solutes.  We have written $\Delta_M^\alpha S$ for the change of entropy of the system $M$ and $\Delta_M^\alpha H$ for the change of enthalpy (enthalpy is energy + a $pressure\times volume$ term).  Similarly, $\Delta_*^\alpha S$ and $\Delta_*^\alpha H$ are the changes for the solutes.]
\medskip
	One can show from the above formulas that the probabilities $\bar p^\alpha$ corresponding to the reverse transitions $(K,\alpha^{\rm out})\to(J,\alpha^{\rm in})$ satisfy
$$	\bar p^\alpha={p^\alpha\exp[\beta(\Delta_M^\alpha+\Delta_*^\alpha)]
	\over\sum_\gamma p^\gamma\exp[\beta(\Delta_M^\gamma+\Delta_*^\gamma)]}\eqno{(5)}   $$
(see [8], Remark 2(d)).
\medskip
	If $K$ is a stationary state, so that $\pi_\tau(K\to K)=1$ we obtain $\sum_\alpha p^\alpha\exp[\beta(\Delta_M^\alpha+\Delta_*^\alpha)]=1$ from (4).  Then (5) gives
$$	\bar p^\alpha=p^\alpha\exp[\beta(\Delta_M^\alpha+\Delta_*^\alpha)]   $$
From this one can deduce that in the average $\Delta_*^\alpha$ is negative: the Gibbs free energy of solutes decreases and is dissipated in the bath, as discussed below in more details.
\bigskip\noindent
{\bf 3. Characterization of pre-biological states.}
\medskip
	We have discussed pre-biological states $J,K,\cdots$ of a system $M$ in contact with a bath containing solutes.  What does non-equilibrium statistical mechanics say about such states?  A good point to start is stationary states: $\pi_\tau(J\to K)=\pi_\tau(K\to L)=\cdots=1$, which means that the growth of the state is limited by the environment.  To qualify as pre-biological, the state $K$ should contain a fair amount of Gibbs free energy.  As seen in Section 2, the stationarity of $K$ implies that
$$	\sum_\alpha p^\alpha\exp[\beta(\Delta_M^\alpha+\Delta_*^\alpha)]=1   $$
and the time-reversal invariance gives
$$	\bar p^\alpha=p^\alpha\exp[\beta(\Delta_M^\alpha+\Delta_*^\alpha)]\eqno{(6)}   $$
so that
$$	p^\alpha=(p^\alpha+\bar p^\alpha){1\over1+\exp[\beta(\Delta_M^\alpha+\Delta_*^\alpha)]}\quad,\quad
	\bar p^\alpha=(p^\alpha+\bar p^\alpha)
	{\exp[\beta(\Delta_M^\alpha+\Delta_*^\alpha)]\over1+\exp[\beta(\Delta_M^\alpha+\Delta_*^\alpha)]}   $$
Writing $u^\alpha=-\beta(\Delta_M^\alpha+\Delta_*^\alpha)$ we have
$$	-\beta\sum_\alpha p^\alpha(\Delta_M^\alpha+\Delta_*^\alpha)
	=\sum_{\alpha:u^\alpha>0}(p^\alpha-\bar p^\alpha)u^\alpha
	=\sum_{\alpha:u^\alpha>0}(p^\alpha+\bar p^\alpha){1-\exp(-u^\alpha)\over1+\exp(-u^\alpha)}u^\alpha   $$
$$	=\sum_{\alpha:u^\alpha>0}(p^\alpha+\bar p^\alpha)u^\alpha\tanh(u^\alpha/2)\ge0   $$
For a stationary state $\sum_\alpha p^\alpha\Delta_M^\alpha=0$ so that the average Gibbs free energy (=enthalpy) released to the bath is $\sum_\alpha p^\alpha(-\Delta_*^\alpha)>0$ (or =0 only if $u^\alpha=0$ identically).  We take the volume of $M$ considered and the time $\tau$ sufficiently small so that there are visible fluctuations of $u^\alpha$.  In particular, since the state $K$ contains a fair amount of Gibbs free energy, there are positive as well as negative fluctuations of $\Delta_M^\alpha$.  The probability of the channel $\alpha$ such that $\Delta_M^\alpha>0$ becomes important only if $|\Delta_*^\alpha|-\Delta_M^\alpha$ is sufficiently large (because of the factor $\tanh(u^\alpha/2)$).  Therefore in a given environment the preservation of a pre-biological stationary state will require either small $\Delta_M^\alpha$, or a machinery capable of converting a complex $\alpha^{\rm in}$ into $\Delta_M^\alpha+\alpha^{\rm out}$.  The existence of such complex reactions is the first hurdle to overcome in the creation of a pre-biological state.  Note that a complex chain $\alpha$ of independent Markovian reactions can be arranged with $\Delta_M^\alpha>0$.  The rates of these reactions must however depend on the environment in an uncontrolled manner.  The relation (6) between the probabilities of $\alpha$ and its inverse $-\alpha$ is a specific consequence of time-reversal invariance.
\medskip
	Let the above machinery $\alpha$ correspond to a state $K$ so that $\pi_\tau(K\to K)=1$.  The situation where $K$ is a statistical superposition of $K_1$ and $K_2$ such that $\pi_\tau(K_1\to K_1)=\pi_\tau(K_2\to K_2)=1$ is unstable in general: mixing the components of $K_1$, $K_2$ will produce a state $\tilde K$ with average $\Delta_M^\alpha$ either greater or smaller, and $K$ will be dynamically replaced by $\tilde K$, $K_1$, $K_2$, or destroyed.  Therefore a pre-biological state $K$ is not normally a mixture.  [We ignore here statistical superpositions of states with unrelated chemistry (see Section 1. (d))].  Note that $K$ may have different aspects depending on position with respect to the bath and its nutrients.  In brief, a pre-biological state is generally indecomposable.  This means in particular that the fluctuations in its composition are not large.  The pre-biological state is also stable under small perturbations, except when those lead to a new metabolic pathway, changing the nature of the state.
\medskip
	We see thus a pre-biological system as a set of components undergoing an organized set of chemical reactions using a limited amount of nutrients in the surrounding fluid.  The complexity of the pre-biological system increases as the amount of available nutrients decreases.  The system can sustain a limited amount of disturbance.  An excessive level of disturbance destroys the organized set of reactions on which the system is based: it dies.
\vfill\eject
\noindent
{\bf References.}
\medskip\noindent
[1] D. Andrieux and P. Gaspard.  ``Nonequilibrium generation of information in copolymerization processes.''  Proc. Natl. Acad. Sci. USA. {\bf 105},9516-9521(2008).
\medskip\noindent
[2] J.M. Diaz.  {\it Inorganic polyphosphate in the marine environment.}  PhD Thesis Georgia Tech., School of Earth and Atm. Sci., May 2011.
\medskip\noindent
[3] F.J. Dyson.  {\it Origins of Life.}  Cambridge U.P., 1999.
\medskip\noindent
[4] M. Eigen.  ``Selforganization of matter and the evolution of biological macromolecules''.  Die Naturwissenschaften.  {\bf 58},465-523(1971).
\medskip\noindent
[5] J.L. England.  ``Statistical physics of self-replication.''  J. Chem. Phys. {\bf 139},121923 (2013).
\medskip\noindent
[6] R.F. Fox.  ``Contributions to the theory of thermostatted systems II: Least dissipation of Helmholtz free energy in nano-biology.''  arXiv:1503.03350.
\medskip\noindent
[7] G. Gallavotti and E.G.D. Cohen.  "Dynamical ensembles in nonequilibrium
statistical mechanics."  Phys. Rev. Letters {\bf 74},2694-2697(1995).
\medskip\noindent
[8] D. Ruelle.  ``A generalized detailed balance relation.''  J. Statist. Phys, {\bf 164},463-471(2016).  [arXiv:1510.08357 (revised v2)]
\medskip\noindent
[9] E. Schr\"odinger.  {\it What Is Life? : The Physical Aspect of the Living Cell.}  Cambridge U.P., 1944.

\end